\documentclass[aps,twocolumn,showpacs,nofootinbib]{revtex4-1}

\usepackage{amsmath,amssymb,bm}
\usepackage[dvipdfmx]{graphicx}
\usepackage{hyperref}
\usepackage{yfonts}
\usepackage{color}
\newcommand{\dd}{\text{d}}
\newcommand{\U}{\text{U}}

\begin{document}

\title{New dynamic critical phenomena in nuclear and quark superfluids}

\author{Noriyuki Sogabe and Naoki Yamamoto}
\affiliation{Department of Physics, Keio University, Yokohama 223-8522, Japan}

\begin{abstract}
We study the dynamic critical phenomena near the possible high-density QCD critical point inside 
the superfluid phase of nuclear and quark matter. We find that this critical point belongs to a new 
dynamic universality class beyond the conventional classification by Hohenberg and Halperin. 
We show that the speed of the superfluid phonon vanishes at the critical point and that the dynamic 
critical index is $z \approx 2$.  
\end{abstract}
\maketitle

\section{Introduction}
\begin{figure}[t]
\begin{center}
\includegraphics[width=7cm]{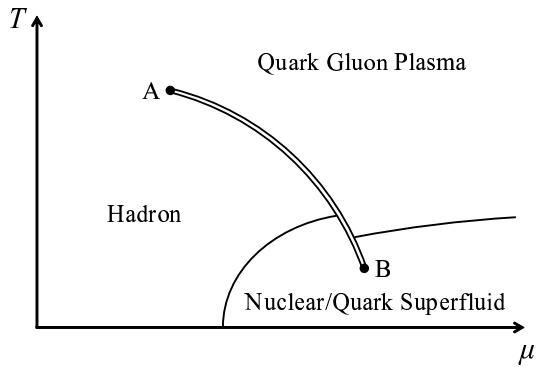}
\end{center}
\caption{A schematic QCD phase structure with the high-temperature critical point A and 
high-density critical point B \cite{Hatsuda:2006ps, Abuki:2010jq}. The double and single lines denote 
the first-order transition associated with chiral symmetry breaking/restoration and the 
second-order transition associated with $\U(1)_{\rm B}$  symmetry breaking/restoration, respectively.}
\label{fig:phase}
\end{figure}

Understanding the phase structure at finite temperature $T$ and baryon chemical potential $\mu$ 
is one of the important problems in quantum chromodynamics (QCD) \cite{Fukushima:2010bq}. 
Among others, it is an open issue to locate the landmark of the QCD phase diagram---the QCD 
critical point \cite{Stephanov:2004wx}. Theoretical analyses suggest the possible existence of the 
high-temperature critical point \cite{Asakawa:1989bq, Barducci:1989wi, Halasz:1998qr, Berges:1998rc, Hatta:2002sj} 
between the hadron and quark-gluon plasma phases as well as high-density critical point 
\cite{Hatsuda:2006ps, Abuki:2010jq, Schmitt:2010pf} (see also Ref.~\cite{Kitazawa:2002bc})
between the hadron and color superconducting phases, as schematically illustrated in 
Fig.~\ref{fig:phase}.%
\footnote{In Refs.~\cite{Hatsuda:2006ps, Abuki:2010jq, Schmitt:2010pf}, the existence of the possible 
high-density critical point was found between the hadron phase and the so-called color-flavor locking 
phase \cite{Alford:1998mk}. In our discussion below, however, what is essential is the fact that 
high-density critical point is inside the superfluid phase with $\U(1)$ baryon number symmetry breaking 
(rather than it is inside the color superconducting phase); even if the chiral critical point exists inside 
the superfluid nuclear matter, our argument is applicable to it. For this reason, high-density region is 
described as the ``nuclear/quark superfluid" in Fig.~\ref{fig:phase}.}
However, not only their locations, but even their existences have not been 
established in QCD {\it per se}. This is mainly because first-principles lattice QCD calculations 
based on Monte Carlo sampling are not feasible due to the so-called sign problem. Therefore, it 
is important to provide theoretical predictions for their critical phenomena to be potentially tested 
in the beam energy scan program at the Relativistic Heavy Ion Collider (RHIC) and future experiments at the
Facility of Antiproton and Ion Research (FAIR), Nuclotron-based Ion Collider Facility (NICA), and 
Japan Proton Accelerator Research Complex (J-PARC).

Generally, critical phenomena around critical points do not depend on the microscopic details of 
the systems, but only on the symmetries and low-energy degrees of freedom. This feature allows 
us to make model-independent predictions for critical phenomena. So far, theoretical studies of 
the QCD critical phenomena have been limited to the high-temperature critical point. It has been 
argued that its static universality class is that of the 3D Ising model \cite{Stephanov:2004wx}, 
and the dynamic universality class is the so-called model H \cite{Fujii:2003bz, Fujii:2004jt, Son:2004iv, Minami:2011un} 
in the classification by Hohenberg and Halperin \cite{Hohenberg:1977ym}. On the other hand, the 
universality class of the high-density QCD critical point has not yet been elucidated until now. 

In this paper, we study the static and dynamic universality classes of the high-density QCD critical 
point.%
\footnote{Here, we will not consider the high-density critical point in the two-flavor color 
superconductivity in Ref.~\cite{Kitazawa:2002bc}. Because $\U(1)_{\rm B}$ symmetry is {\it not} 
broken in this phase \cite{Alford:2007xm}, both of the static and dynamic universality classes are 
the same as those of the high-temperature critical point. Our focus in this paper will be on the 
high-density critical point with $\U(1)_{\rm B}$ symmetry breaking in 
Refs.~\cite{Hatsuda:2006ps, Abuki:2010jq} (see also footnote 1).}
We show that its static universality class is the same as that of the high-temperature QCD critical 
point. On the other hand, we find that its dynamic universality class is different from not only that 
of the high-temperature QCD critical point, but also those of all the models in the classification by 
Hohenberg and Halperin \cite{Hohenberg:1977ym}; to the best of our knowledge, this is a {\it new} 
dynamic universality class that has not been found in any other system. As we will discuss in this 
paper, its uniqueness stems from the interplay between the chiral criticality and the presence of 
the superfluid phonon---a feature specific for the high-density QCD critical point.
In other words, experimental identification of this unique dynamic critical phenomenon would 
provide indirect evidence of the superfluidity in the high-density regime of QCD.

This paper is organized as follows. In Sec.~\ref{sec:hydro}, we identify a set of hydrodynamic 
variables near the high-density QCD critical point. In Secs.~\ref{sec:statics} and \ref{sec:dynamics}, 
we study the static and dynamic critical phenomena near the high-density QCD critical point, 
respectively, by using the simplified model without the energy-momentum density. 
In Sec.~\ref{sec:EM} we consider the full hydrodynamic modes and show that both the static 
and dynamic universality classes are the same as those in Secs.~\ref{sec:statics} and \ref{sec:dynamics}. 
Finally, we conclude with Sec.~\ref{sec:conclusion}, with the discussion on the physical reason 
why the high-density QCD critical point belong to the new dynamic universality class. 

\section{Hydrodynamic variables}
\label{sec:hydro}
In general, critical phenomena near critical points at long distance and long time scale are 
independent of microscopic details and depend only on the low-energy degrees of freedom. To 
describe the critical phenomena of a given system, it is thus sufficient to consider the low-energy 
effective theory for slow degrees of freedom called the hydrodynamic variables. Such a 
low-energy effective theory can be written down based on the systematic derivative expansion 
guided by the symmetries of the system. Typically, the hydrodynamic variables are the 
fluctuations of the conserved charges, the order parameter associated with the critical phenomena, 
and the Nambu-Goldstone modes associated with spontaneous breaking of some symmetries.

Near the high-density QCD critical point, the hydrodynamic variables are the fluctuations of the 
following quantities:
\begin{enumerate}
\item{The conserved energy and momentum densities,  
$\varepsilon \equiv T^{00}-\langle T^{00}\rangle$ and $\pi^{i} \equiv T^{0i}$;}
\item{The conserved baryon number density, $n \equiv \bar q \gamma^0 q - \langle \bar q \gamma^0 q \rangle$;}
\item{The chiral condensate, $\sigma \equiv \bar q q - \langle \bar q q \rangle$;}
\item{The superfluid phonon $\theta$ associated with the spontaneous breaking 
of the $\U(1)_{\rm B}$ symmetry, $q \rightarrow q e^{i \theta}$.}
\end{enumerate}
The hydrodynamic variables (i)--(iii) above are the same as those of the high-temperature QCD 
critical point \cite{Son:2004iv}, but here we have the additional hydrodynamic variable (iv) due 
to the nuclear/quark superfluids. 

Note that the fluctuation of the amplitude of the diquark condensate (which we denote by $\phi$)
is not the hydrodynamic variable, because the high-density QCD critical point here is 
characterized by massless $\sigma$, and not by $\phi$; the high-density QCD critical point is not 
related to the second-order superfluid phase transition where $\phi$ becomes massless (see 
Fig.~\ref{fig:phase}). Note also that Nambu-Goldstone modes associated with chiral symmetry 
breaking do not enter our low-energy effective theory because they acquire finite mass $m_{\pi}$ 
in the presence of finite quark masses; we are interested in the low-energy physics with the length 
scale much larger than $1/m_{\pi}$. In the color superconducting phase, the gluons acquire the 
mass gap due to the color Meissner effect and do not enter the low-energy effective theory as well.

In the following, we will construct the low-energy effective theory describing the static and 
dynamic critical phenomena near the high-density QCD critical point, based on the symmetries of 
QCD at finite $T$ and $\mu$: $\U(1)_{\rm B}$ symmetry, chiral symmetry, and discrete ${\cal CPT}$ 
symmetries (charge conjugation, parity, and time reversal symmetry). Note that the time reversal 
symmetry is macroscopically broken by the presence of dissipation in the dynamical case.

As it will turn out that the energy and momentum densities, $\varepsilon$ and $\pi^i$, do not 
affect the static and dynamic critical phenomena, we will first consider the case only with the 
hydrodynamic variables $x_{i} \equiv \sigma, n, \theta$. We will then consider the case also with 
$\varepsilon$ and $\pi^i$ in Sec.~\ref{sec:EM} and show that they do not actually affect the static 
and dynamic universality classes.

\section{Static universality class}
\label{sec:statics}
Let us first study the static properties of the critical phenomena near the high-density QCD 
critical point. The general Ginzburg-Landau potential consistent with the QCD symmetries (with 
finite quark mass $m_{\rm q}$ and at finite chemical potential $\mu$) to the second order is given by
\begin{align}
\label{eq:GL}
F[\displaystyle \sigma,n,\theta]
=\int \dd \bm{r} \biggl[
\frac{a}{2}({\bm \nabla}\sigma)^{2}+b{\bm \nabla}\sigma \cdot {\bm \nabla}n+\frac{c}{2}({\bm \nabla}n)^{2} 
\nonumber \\
 +\frac{d}{2}({\bm \nabla}\theta)^{2}+V(\sigma,n ) \biggr] \,, 
\end{align}
where
\begin{align}
\label{eq:V}
V(\displaystyle \sigma,n)=\frac{A}{2}\sigma^{2}+B\sigma n+\frac{C}{2}n^{2}\,.
\end{align}
Here the coefficients $a,b,c,d,A,B,C$ are the functions of $T$ and $\mu$ that depend on the 
microscopic details. Importantly, the mixing between $\sigma$ and $n$ is allowed in the presence 
of $m_{\rm q}$ and $\mu$ \cite{Son:2004iv}. On the other hand, the mixing terms between 
$\theta$ and $\sigma$ or $n$ are prohibited by the time reversal symmetry in this static 
Ginzburg-Landau potential. (Note here that $\theta$ is ${\cal T}$ odd.)

One can calculate the correlation functions and static responses to external perturbations by
\begin{align}
\label{eq:expect}\displaystyle \left\langle\mathcal{O}[x_{j}]\right\rangle
=\frac{\displaystyle\int\prod_{i} \mathcal{D}x_{i}\mathcal{O}[x_{j}]e^{-\beta F}}{\displaystyle\int\prod_{i} \mathcal{D}x_{i}e^{-\beta F}}\,.
\end{align}
Because $\theta$ is decoupled from $\sigma$ and $n$, the correlation function of $\sigma$ and 
the baryon number susceptibility defined by $\chi_{\rm B}\equiv {\delta n}/{\delta \mu}$ are the 
same as those for the high-temperature QCD critical point obtained in Ref.~\cite{Son:2004iv}:
\begin{gather}
\label{eq:corr}
\displaystyle \left\langle\sigma(\bm{r})\sigma({\bm 0})\right\rangle=\frac{1}{4\pi r} e^{-r/\xi}\,, \\
\label{eq:chi}
\chi_{\rm B} =\frac{1}{\mathcal{V} T}\left\langle n^{2}\right\rangle_{\bm{q}\rightarrow {\bm 0}}=\frac{A}{\Delta}\,,
\end{gather}
where $\mathcal{V}$ denotes the spatial volume and
\begin{align}
\label{eq:crit}
\xi\sim\Delta^{-\frac{1}{2}}, \qquad \Delta\equiv AC-B^{2}.
\end{align}

The critical point is characterized by the condition that the correlation length diverges, 
$\xi\rightarrow\infty$, or $\Delta\rightarrow0$. Due to the mixing between $\sigma$ and $n$, 
there is only one linear combination of $\sigma$ and $n$ that becomes massless near the 
critical point \cite{Son:2004iv}. Note here that $\theta$ is irrelevant to the condition for the 
criticality. Therefore, the {\it static} universality class of the high-density QCD critical point is 
the same as that of the high-temperature one obtained in Ref.~\cite{Son:2004iv}: the universality 
class of the 3D Ising model. Then, the critical exponent of $\chi_{\rm B}$ for the high-density 
QCD critical point can be obtained as 
\begin{align}
\label{eq:chiB}
\chi_{\rm B}\sim\xi^{2-\eta}
\end{align}
with $\eta\simeq 0.04$.

\section{Dynamic universality class}
\label{sec:dynamics}

\subsection{Langevin equation}
\label{sec:Langevin}
We now discuss the dynamics of hydrodynamic variables near the high-density QCD critical 
point using the low-energy effective theory---the Langevin equation. Following the standard 
procedure (see, e.g., Ref.~\cite{Chaikin}), one can write down the generic Langevin theory for 
the hydrodynamic variables $x_i = \sigma, n, \theta$ as 
\begin{align}
\label{eq:Langesigma}
\displaystyle\dot{\sigma}(\bm{r})&=-\Gamma\frac{\delta F}{\delta\sigma(\bm{r})}+\tilde{\lambda}\bm{\nabla}^{2}\frac{\delta F}{\delta n(\bm{r})}
\nonumber \\
& \quad -\int \displaystyle \dd \bm{r}'[\tilde{\sigma}(\bm{r}),\theta(\bm{r}')]\frac{\delta F}{\delta\theta(\bm{r}')}+\xi_{\sigma}(\bm{r})\,,
\\
\label{eq:Langen}
\displaystyle\dot{n}(\bm{r})&=\tilde{\lambda}\bm{\nabla}^{2}\frac{\delta F}{\delta\sigma(\bm{r})}+\lambda\bm{\nabla}^{2}\frac{\delta F}{\delta n(\bm{r})}
\nonumber \\
& \quad -\int \displaystyle \dd \bm{r}'[\tilde{n}(\bm{r}),\theta(\bm{r}')]\frac{\delta F}{\delta\theta(\bm{r}')}+\xi_{n}(\bm{r})\,,
\\
\label{eq:Langetheta}
\displaystyle\dot{\theta}(\bm{r})&=- \int \dd \bm{r}'[\theta(\bm{r}),\tilde{\sigma}(\bm{r}')]\frac{\delta F}{\delta\sigma(\bm{r}')}
\nonumber \\
& \quad - \int \dd \bm{r}'[\theta(\bm{r}),\tilde{n}(\bm{r}')]\frac{\delta F}{\delta n(\bm{r}')}-\zeta\frac{\delta F}{\delta\theta(\bm{r})}+\xi_{\theta}(\bm{r})\,.
\end{align}
Here $\Gamma$, $\lambda$, and $\tilde \lambda$ are the kinetic coefficients related to dissipation, 
$\xi_{\sigma}$, $\xi_n$, and $\xi_{\theta}$ are the noise terms, and we defined $\tilde{\sigma} \equiv \bar q q$ 
and $\tilde{n} \equiv \bar q \gamma^0 q$. [On the other hand, note again that $\sigma$ and $n$ are defined 
as the fluctuations around the expectation values, $\sigma(\bm{r}) = \tilde{\sigma}(\bm{r})-\sigma_{\text{eq}}$ 
and $n(\bm{r})=\tilde{n}(\bm{r})-n_{\text{eq}}$.] 
The angler brackets denote the Poisson brackets, which we postulate as 
\begin{align}
\label{eq:sigtheta}
[\tilde{\sigma}(\bm{r}),\theta(\bm{r}')]
&=0\,, \\
\label{eq:ntheta}
[\theta(\bm{r}),\tilde{n}(\bm{r}')]&=\delta(\bm{r}-\bm{r}')\,.
\end{align} 

Equation~(\ref{eq:ntheta}) shows that $n$ is canonical conjugate to $\theta$. This can be understood 
from the $\U(1)$ gauge symmetry, which dictates that the Lagrangian for the superfluid phonon 
depends on $\theta$ through the form ${\cal L} = {\cal L}(\dot \theta + \mu, {\bm \nabla} \theta)$
\cite{Weinberg:1996kr, Son:2002zn}, so that 
\begin{align}
\tilde n \equiv \frac{\delta {\cal L}}{\delta \mu} = \frac{\delta {\cal L}}{\delta \dot \theta}\,. \nonumber
\end{align}

Some remarks on the Langevin equations above are in order here. The coefficients of the first 
two terms on the right-hand side of Eq.~(\ref{eq:Langen}) include derivatives to satisfy the baryon 
number conservation, $\dot n = 0$, in the long wavelength limit, ${\bm q} \rightarrow {\bm 0}$. 
According to Onsager's principle, the coefficient of the second term in Eq.~(\ref{eq:Langesigma}) 
must be the same as that of the first term in Eq.~(\ref{eq:Langen}). The correlators of 
$\xi_{\sigma}$, $\xi_n$, and $\xi_{\theta}$ are chosen such that the equilibrium distribution 
$e^{- \beta F}$ is reproduced. Because the noise terms are not important at the mean-field level, 
we will ignore them in the following. 

The variations $\delta F/ \delta x_i$ can be computed by using the Ginzburg-Landau potential
(\ref{eq:GL}) as
\begin{align}
\label{eq:sigmaconj}
\displaystyle \frac{\delta F}{\delta\sigma}&=(A-a\bm{\nabla}^{2})\sigma+(B-b\bm{\nabla}^{2})n\,, \\
\displaystyle \frac{\delta F}{\delta n}&=(B-b\bm{\nabla}^{2})\sigma+(C-c\bm{\nabla}^{2})n\,, \\
\label{eq:thetaconj}
\displaystyle \frac{\delta F}{\delta\theta}&=-d\bm{\nabla}^{2}\theta\,.
\end{align}
 
By substituting Eqs.~(\ref{eq:sigtheta})--(\ref{eq:thetaconj}) into Eqs.~(\ref{eq:Langesigma})--(\ref{eq:Langetheta}) 
in momentum space $(\omega, {\bm q})$, the Langevin equation to the order of $O({\bm q}^2)$ is 
summarized in the matrix form,
\begin{widetext}
\begin{align}
\label{eq:Mx=0}
\mathcal{M}\bm{x}\equiv\left(\begin{array}{ccc}
i\omega-\Gamma A-(\Gamma a+\tilde{\lambda}B)\bm{q}^{2} &-\Gamma B-(\Gamma b+\tilde{\lambda}C)\bm{q}^{2}&0\\
-(\tilde{\lambda}A+\lambda B)\bm{q}^{2}& i\omega-(\tilde{\lambda}B+\lambda C)\bm{q}^{2} & d\bm{q}^{2}\\
-B-b\bm{q}^{2} &-C-c\bm{q}^{2} &i\omega-\zeta d\bm{q}^{2}
\end{array}\right)\left(\begin{array}{c}
\sigma\\
n\\
\theta
\end{array}\right)=0\,.
\end{align}
\end{widetext}

\subsection{Hydrodynamic modes}
\label{sec:mode}
The hydrodynamic modes can be obtained by solving the proper equation, $\det \mathcal{M}=0$. 
This equation reduces to 
\begin{align}
\label{eq:prop}
\omega^{3}+i(x_{1}+x_{2}\bm{q}^{2})\omega^{2}-y\bm{q}^{2}\omega-iz\bm{q}^{2}=0\,,
\end{align}
where
\begin{align}
x_{1}&\equiv\Gamma A\,, \nonumber \\
x_{2}&\equiv\Gamma a+2\tilde{\lambda}B+\lambda C+\zeta d\,, \nonumber \\
y&\equiv\Gamma\lambda\Delta+Cd+\Gamma\zeta Ad\,, \nonumber \\
z&\equiv\Gamma d\Delta\,.
\end{align}
In Eq.~(\ref{eq:prop}), we ignore the higher-order terms in ${\bm q}$. At this order, the left-hand 
side of Eq.~(\ref{eq:prop}) can be factorized as
\begin{align}
& \ \ \ \left[\omega+ix_{1}+i\left(x_{2}-\frac{y}{x_{1}}+\frac{z}{x_{1}^{2}}\right)\bm{q}^{2} + O(\bm{q}^{3}) \right] 
\nonumber \\
& \times \left[\omega-\sqrt{\frac{z}{x_{1}}}|\bm{q}|+\frac{i}{2}\left(\frac{y}{x_{1}}-\frac{z}{x_{1}^{2}}\right)\bm{q}^{2}
+ O(\bm{q}^{3}) \right]
\nonumber \\
& \times \left[\omega+\sqrt{\frac{z}{x_{1}}}|\bm{q}|+\frac{i}{2}\left(\frac{y}{x_{1}}-\frac{z}{x_{1}^{2}}\right)\bm{q}^{2}
+ O(\bm{q}^{3}) \right]=0\,.
\end{align}
This shows that the system near the high-density QCD critical point has three hydrodynamic 
modes: the relaxation mode and the pair of phonons with the dispersion relations,
\begin{align}
\omega_{1}&=-i\Gamma A+O(\bm{q}^{2})\,,\\
\label{eq:omega23}
\omega_{2,3}&=\pm c_{\rm s}|\bm{q}|+O(\bm{q}^{2})\,,
\end{align}
respectively. Here
\begin{align}
\label{eq:cs}
c_{\rm s}\equiv \displaystyle\sqrt{\frac{d}{\chi_{\rm B}}}
\end{align}
is the speed of the phonon.

We find that $c_{\rm s} \rightarrow 0$ due to the divergence of $\chi_{\rm B}$ when the critical point 
is approached [see Eq.~(\ref{eq:chiB})]. This is the phenomenon of the critical slowing down.

\subsection{Dynamic critical exponent}
\label{sec:z}
The dynamic critical phenomena can be characterized by the dynamic critical exponent $z$ 
defined by $\tau \sim \xi^{z}$, or
\begin{align}
\label{eq:zdef}
c_{\rm s}\sim\xi^{1-z}\,.
\end{align}

To find the value of $z$, we use the following $\xi$ dependences of $d$ and $\chi_{\rm B}$ 
near the high-density QCD critical point:
\begin{align}
\label{eq:paraxi}
d\sim\xi^{0}, \quad \chi_{\rm B}\sim\xi^{2-\eta}.
\end{align}
$d$ is the stiffness parameter (or the ``decay constant" for the superfluid phonon), and it does 
not depend on $\xi$ close to the high-density critical point away from the superfluid phase 
transition. The behavior of $\chi_{\rm B}$ near the critical point is given by Eq.~(\ref{eq:chiB}). 

From Eqs.~(\ref{eq:cs}), (\ref{eq:zdef}), and (\ref{eq:paraxi}), one finds the dynamic critical 
exponent $z$ as
\begin{align}
\label{eq:z}
z = 2 - \frac{\eta}{2}\,.
\end{align}
This is different from those of the other models in the classification by Hohenberg and Halperin 
\cite{Hohenberg:1977ym}, indicating that the high-density QCD critical point belongs to a new 
dynamic universality class.

\section{Coupling to energy and momentum densities}
\label{sec:EM}
So far we have ignored the contributions of the energy and momentum densities, $\varepsilon$ 
and $\bm{\pi}$. In this section, we show that these contributions do not affect the static and 
dynamic universality classes of the high-density QCD critical point obtained above.   

\subsection{Statics}\label{sec:EMsta}
Let us first consider the Ginzburg-Landau functional to the second order in terms of $\sigma$, $n$, 
$\varepsilon$, $\bm{\pi}$, and $\theta$ describing static critical phenomena. Similarly to the 
argument in Sec.~\ref{sec:statics}, the time reversal symmetry prohibits the mixing between 
$x_{i}\equiv\sigma,n, \varepsilon$ and $\bm{\pi},\theta$:
\begin{align}
F[\sigma,n,\varepsilon,\bm{\pi},\theta]&=F[\sigma,n,\varepsilon]+F[\bm{\pi},\theta]\,.
\end{align}
The presence of $m_{\rm q}$ and $\mu$ generally allows the mixing among $\sigma,\ n,$ and 
$\varepsilon$ in the $\mathcal{T}$-even sector as
\begin{gather}
\label{F_even}
F[\displaystyle \sigma,n,\varepsilon]=\frac{1}{2}\int \dd \bm{r} x_{i}\beta_{ij}(\bm{\nabla})x_{j}\,, \\
\beta_{ij}(\bm{\nabla})=V_{ij}-v_{ij}\bm{\nabla}^{2}\,.
\end{gather}
Here the subscripts $i, j$ are the shorthand notations for $x_{i}, x_{j}$. The $\mathcal{T}$-odd 
sector is given by
\begin{gather}
\label{F_odd}
F[\displaystyle \bm{\pi},\theta]=\frac{1}{2}\int \dd \bm{r}\left[V_{\pi\pi}\bm{\pi}^{2}+2V_{\pi\theta}\bm{\pi}\cdot\bm{\nabla}\theta+V_{\theta\theta}(\bm{\nabla}\theta)^{2}\right]\,.
\end{gather}
We call $V_{ij}$,\ $v_{ij}$, $V_{\pi\pi}$, $V_{\pi\theta}$, and $V_{\theta\theta}$ the 
Ginzburg-Landau parameters. 

By completing the square, Eq.~(\ref{F_odd}) becomes
\begin{align}
F[\displaystyle \bm{\pi},\theta]=\frac{1}{2}\int \dd \bm{r}\left[V_{\pi\pi}({\bm \pi}')^{2}+V_{\theta\theta}'(\bm{\nabla}\theta)^{2}\right]\,,
\end{align}
where ${\bm \pi}' \equiv \bm{\pi}+V_{\pi\theta} \bm{\nabla}\theta/V_{\pi\pi}$ and 
$V_{\theta\theta}' \equiv V_{\theta\theta}-V_{\pi\theta}^{2}/V_{\pi\pi}$. Below we assume that
$V_{\theta\theta}'>0$ and redefine ${\bm \pi}'$ and $V_{\theta\theta}'$ as ${\bm \pi}$ and 
$V_{\theta\theta}$ for simplicity. 

In a way similar to Eq.~(\ref{eq:corr}) in Sec.~\ref{sec:statics}, the correlation length $\xi$ is 
defined from the correlation function of $\sigma$. In the present case, we find
\begin{align}
\xi\sim(\det V)^{-\frac{1}{2}}\,.
\end{align}
The critical point is characterized by the condition, $\det V=0$, analogously to the condition 
$\Delta = 0$ in Sec.~\ref{sec:statics}. At the critical point, only one of the linear combinations of 
$\sigma$, $n$, and $\varepsilon$ becomes massless. Due to the degeneracy among $\sigma$, 
$n$, and $\varepsilon$, the static universality class remains the same as that of Sec.~\ref{sec:statics}.

We define the generalized susceptibilities 
\begin{align}
\displaystyle \chi_{ij} \equiv \left. \frac{\delta\langle x_{i}\rangle_{X_j}}{\delta X_{j}} \right|_{X_{j}=0} \,,
\end{align}
where $\langle x_{i}\rangle_{X_j}$ is the the expectation value of $x_{i}$ in Eq.~(\ref{eq:expect}) 
with the replacement $\beta F \rightarrow \beta F+\displaystyle \int \dd \bm{r}x_{j}X_{j}$ 
(where summation over the index $j$ is {\it not} implied) with $X_{\varepsilon} \equiv -\beta$, 
$X_{n} \equiv \beta \mu$, and $X_{\sigma} \equiv \beta m_{q}$.
One can show the general relation between $\displaystyle \chi_{ij}$ and $V_{ij}$ as
\begin{align}
\label{eq:chiepep}\chi_{\varepsilon\varepsilon}&=\frac{1}{\mathcal{V}}\langle {\varepsilon}^{2}\rangle_{\bm{q}\rightarrow {\bm 0}}=T(V^{-1})_{\varepsilon\varepsilon}\,,\\
\chi_{\varepsilon n}&=\frac{1}{\mathcal{V}}\langle \varepsilon n \rangle_{\bm{q}\rightarrow {\bm 0}}=T(V^{-1})_{\varepsilon n}\,,\\
\label{eq:chinn}
\chi_{nn}&=\frac{1}{\mathcal{V}}\langle n^{2}\rangle_{\bm{q}\rightarrow {\bm 0}}=T(V^{-1})_{nn}\,,
\end{align} 
where $(V^{-1})_{ij}$ denotes the $(i,j)$ component of the inverse matrix of $V$. Since 
$(V^{-1})_{ij}\propto(\det V)^{-1}$, one finds
\begin{align}
\label{eq:chixis}
\chi_{\varepsilon\varepsilon}\sim\chi_{\varepsilon n}\sim\chi_{nn}\sim\xi^{2-\eta}\,,
\end{align}
with $\eta \simeq 0.04$.

\subsection{Dynamics}
\label{sec:fulldynamics}
\subsubsection{Full Langevin equations}
To describe the dynamic critical phenomena, we now consider the Langevin equations in terms 
of the full hydrodynamic variables $x_{i}=\sigma,n, \varepsilon$, and $\bm{\pi},\theta$. The 
Langevin equations read 
\begin{align}
\label{eq:langex}
\displaystyle \dot{x}_{i}(\bm{r})&=-\gamma_{ij}(\bm{r})\frac{\delta F}{\delta x_{j}(\bm{r})}
-\int \dd \bm{r}'\left[\tilde{x}_{i}(\bm{r}),\bm{\pi}(\bm{r}')\right]\cdot\frac{\delta F}{\delta\bm{\pi}(\bm{r}')}\notag \\
&\quad-\displaystyle \int \dd \bm{r}'\left[\tilde{x}_{i}(\bm{r}),\theta(\bm{r}')\right]\frac{\delta F}{\delta\theta(\bm{r}')}+\xi_{i}(\bm{r})\,,\\
\label{eq:langepi}
\displaystyle \dot{\bm{\pi}}(\bm{r})&=\Gamma_{\pi\pi}\bm{\nabla}\bm{\nabla}\cdot\frac{\delta F}{\delta\bm{\pi}(\bm{r})}+\Gamma_{\pi\pi}'\bm{\nabla}^{2}
\frac{\delta F}{\delta\bm{\pi}(\bm{r})}-\Gamma_{\pi\theta}\bm{\nabla}\frac{\delta F}{\delta\theta(\bm{r})}\notag \\
&\quad-\displaystyle \int \dd \bm{r}'\left[\bm{\pi}(\bm{r}),\tilde{x}_{i}(\bm{r}')\right]\frac{\delta F}{\delta x_{i}(\bm{r}')}+\bm{\xi}_{\pi}(\bm{r})\,,\\
\label{eq:langetheta2}
\displaystyle \dot{\theta}(\bm{r})&=-\Gamma_{\pi\theta}\bm{\nabla}\cdot\frac{\delta F}{\delta\bm{\pi}(\bm{r})}-\Gamma_{\theta\theta}\frac{\delta F}{\delta\theta(\bm{r})}\notag \\
&\quad-\displaystyle \int \dd \bm{r}'\left[\theta(\bm{r}),\tilde{x}_{i}(\bm{r}')\right]\frac{\delta F}{\delta x_{i}(\bm{r}')}+\xi_{\theta}(\bm{r})\,,
\end{align}
where summation over repeated indices $i, j$ is understood. Here 
\begin{align}
\gamma_{ij}\equiv\left(\begin{array}{ccc}
\Gamma_{\sigma\sigma} &-\Gamma_{\sigma n}\bm{\nabla}^{2} &-\Gamma_{\sigma \varepsilon}\bm{\nabla}^{2}\\
-\Gamma_{\sigma n}\bm{\nabla}^{2}& -\Gamma_{n n}\bm{\nabla}^{2} &-\Gamma_{n \varepsilon}\bm{\nabla}^{2}\\
-\Gamma_{\sigma \varepsilon}\bm{\nabla}^{2}& -\Gamma_{n \varepsilon}\bm{\nabla}^{2}& -\Gamma_{\varepsilon \varepsilon}\bm{\nabla}^{2}
\end{array}\right)\,,
\end{align}
$\Gamma_{\alpha\beta}$ ($\alpha,\beta=x_{i},\pi,\theta$), and $\Gamma_{\pi\pi}'$ are the kinetic 
coefficients, and $\tilde{\varepsilon} \equiv T^{00}$. The noise terms $\xi_{i},\ \bm{\xi}_{{\pi}},$ and 
$\xi_{\theta}$ above are not important in the following discussion. To write down the equations 
above, we took into account the momentum conservation law and the Onsager's principle, 
similarly to Sec.~\ref{sec:dynamics}.

We postulate the Poisson brackets, in addition to Eqs.~(\ref{eq:sigtheta}) and (\ref{eq:ntheta}), 
as follows:
\begin{align}
\label{eq:brasig}
[\bm{\pi}(\bm{r}),\tilde{\sigma}(\bm{r}')]&=\tilde{\sigma}(\bm{r}')\bm{\nabla}\delta(\bm{r}-\bm{r}')\,,\\
\label{eq:brapin}[\bm{\pi}(\bm{r}),\tilde{n}(\bm{r}')]&=\tilde{n}(\bm{r}')\bm{\nabla}\delta(\bm{r}-\bm{r}')\,, \\
[\bm{\pi}(\bm{r}),\tilde{s}(\bm{r}')]&=\tilde{s}(\bm{r}')\bm{\nabla}\delta(\bm{r}-\bm{r}')\,,\\
\label{eq:brathetas}
[\theta(\bm{r}),\tilde{s}(\bm{r}')]&=0\,.
\end{align}
Here $\tilde{s}(\bm{r})$ denotes the entropy density. The Poisson brackets concerning 
$\tilde{\varepsilon}(\bm{r})$ can be derived as
\begin{align}
\label{eq:brapiepsilon}
[\bm{\pi}(\bm{r}),\tilde{\varepsilon}(\bm{r}')]&=\left(T\tilde{s}(\bm{r})+\mu \tilde{n}(\bm{r})\right) \bm{\nabla}\delta(\bm{r}-\bm{r}')\,,\\
\label{eq:brathetaepsilon}
[\theta(\bm{r}),\tilde{\varepsilon}(\bm{r}')]&=\mu\delta(\bm{r}-\bm{r}')\,.
\end{align}
Here, we used the thermodynamic relation, $\dd \varepsilon=T \dd s+\mu \dd n$ and the 
definition of the Poisson brackets 
$\displaystyle \left[\bm{\pi}(\bm{r}),\tilde{y}_{i}(\bm{r}')\right] \equiv \delta \tilde{y}_{i}(\bm{r}')/\delta \bm{u}(\bm{r})$ 
for $\tilde{y}_{i}\equiv \tilde{n},\tilde{s},\tilde{\varepsilon}$, with $\bm{u}(\bm{r})$ the infinitesimal 
translation of the coordinate, $\bm{r}\rightarrow\bm{r}+\bm{u}(\bm{r})$. 
(For the details of the Poisson brackets, see, e.g., Ref.~\cite{Volovick}.)

We consider the small fluctuations of variables around the equilibrium values, 
$\sigma(\bm{r}) = \tilde{\sigma}(\bm{r})-\sigma_{\text{eq}}$, $n(\bm{r}) = \tilde{n}(\bm{r}) - n_{\text{eq}}$, 
and $s(\bm{r}) = \tilde{s}(\bm{r}) - s_{\text{eq}}$. Then, the linearized Langevin equations can be 
written as follows:
\begin{align}
\label{eq:linearsig}
\dot{\sigma} &= -\left(\Gamma_{\sigma \sigma} V_{\sigma i} - \Gamma_{\sigma j} v_{j i} \bm{\nabla}^{2}\right) x_i 
- \displaystyle \sigma_{\text{eq}}V_{\pi\pi}\bm{\nabla}\cdot\bm{\pi}\,,\\
\dot{n} &= \Gamma_{n j}v_{j i}\bm{\nabla}^{2}x_i 
-\displaystyle n_{\text{eq}}V_{\pi\pi}\bm{\nabla}\cdot\bm{\pi}-\displaystyle V_{\theta\theta}\bm{\nabla}^{2}\theta\,,\\
\dot{\varepsilon} &= \Gamma_{\varepsilon j}v_{j i}\bm{\nabla}^{2}x_i 
-w_{\text{eq}}V_{\theta\theta}\bm{\nabla}\cdot\bm{\pi}-\mu V_{\theta\theta}\bm{\nabla}^{2}\theta\,,\\
\dot{\bm{\pi}} &= -\left(\sigma_{\mathrm{e}\mathrm{q}}V_{\sigma i}+n_{\mathrm{e}\mathrm{q}}V_{n i}+w_{\text{eq}}V_{\varepsilon i}\right)\bm{\nabla}x_i \notag \\
&\quad+\Gamma_{\pi\pi}V_{\pi\pi}\bm{\nabla}\bm{\nabla}\cdot\bm{\pi}
+\Gamma_{\pi\pi}'V_{\pi\pi}\bm{\nabla}^{2}\bm{\pi}\,, \\
\label{eq:lineartheta}
\dot{\theta}& = -\left[\left(V_{n i}+\mu V_{\varepsilon i}\right)-\left(v_{n i}+\mu v_{\varepsilon i}\right)\bm{\nabla}^{2}\right]x_i \notag \\
&\quad-\Gamma_{\pi\theta}V_{\pi\pi}\bm{\nabla}\cdot\bm{\pi}
+\Gamma_{\theta\theta} V_{\theta\theta}\bm{\nabla}^{2}\theta\,,
\end{align}
where $w_{\text{eq}} = Ts_{\text{eq}}+\mu n_{\text{eq}}$. 

\subsubsection{Decomposition of momentum density}
It is convenient to work in the $(t, {\bm q})$ space to decompose the momentum density $\pi^{i}$ 
into the longitudinal and transverse parts with respect to momentum ${\bm q}$, 
\begin{align}
\pi^{i}=\pi_{\rm L}^{i}+\pi_{\rm T}^{i}, \quad \pi^i_{\rm L} = (P_{\rm L})^{ij} \pi^j, 
\quad \pi^i_{\rm T} = (P_{\rm T})^{ij} \pi^j,
\end{align}
where $P_{\rm L,T}$ are the longitudinal and transverse projections defined by
\begin{align}
(P_{\rm L})^{ij} \equiv \frac{q^i q^j}{|{\bm q}|^2}\,, \quad  
(P_{\rm T})^{ij} \equiv \delta_{ij} - \frac{q^i q^j}{|{\bm q}|^2}\,.
\end{align}

One can show that the dynamics of $\pi_{\rm T}^{i}$ is decoupled from the dynamics of the other 
variables as follows. The linearized Langevin equations obtained above can be written down to 
the leading order of ${\bm q}$ as
\begin{align}
\label{eq:sigpi}
\dot x_k &= A_{x_k}(\bm{q}\cdot\bm{\pi}) + f(x_k,\theta)\,, \\
\label{eq:pipi}
\dot{\bm \pi} &= A_{\pi} {\bm q}({\bm q}\cdot{\bm \pi}) 
+ B_{\pi} |{\bm q}|^2 {\bm \pi} + {\bm q} g(x_k)\,,\\
\label{eq:thetapi}
\dot{\theta}&=A_{\theta}(\bm{q}\cdot\bm{\pi})+h(x_{k},\theta)\,,
\end{align}
where $f(x_k,\theta)$, $g(x_k)$, and $h(x_k,\theta)$ denote the terms that may involve $x_k$ and 
$\theta$, but do not $\bm{\pi}$. The coefficients $A_{x_k}$, $A_{\pi}$, $A_{\theta}$, and $B_{\pi}$ 
denote the parameters which depend on the Ginzburg-Landau parameters $V_{ij},\ v_{ij}$, kinetic 
coefficients, and equilibrium values of the thermodynamic quantities. The explicit forms of these 
coefficients are not important for our purpose. Using ${\bm \pi}_{\rm L,T}$, 
Eqs.~(\ref{eq:sigpi})--(\ref{eq:thetapi}) can be rewritten as 
\begin{align}
\dot x_k &= A_{x_k}(\bm{q}\cdot\bm{\pi}_{\rm L}) + f(x_k,\theta)\,, \\
\dot{\bm \pi}_{\rm L} &=  (A_{\pi} + B_{\pi}) |{\bm q}|^2 
{\bm \pi}_{\rm L} + {\bm q} g(x_k)\,, \\
\dot{\bm \pi}_{\rm T} &= B_{\pi} |{\bm q}|^2 {\bm \pi}_{\rm T}\,,\\
\dot{\theta}&=A_{\theta}(\bm{q}\cdot\bm{\pi}_{\text{L}})+h(x_{k},\theta)\,.
\end{align}
So the dynamics of $\pi^{i}_{\rm L}$ and $\pi_{\rm T}^{i}$ are decoupled from each other at the 
mean-field level.

\subsubsection{Hydrodynamic modes}
It is easy to obtain the Langevin equation for $\bm{\pi}_{T}$ in the $(\omega, {\bm q})$ space as
\begin{align}
\label{eq:pitdynamics}
\left(i\omega-B_{\pi}\bm{q}^{2}\right)\bm{\pi}_{\mathrm{T}}=0,\quad B_{\pi}=\Gamma_{\pi\pi}'V_{\pi\pi},
\end{align}
which shows the diffusion mode.

The other Langevin equations which involve ${\bm \pi}_{\rm L}$ (but not $\bm{\pi}_{\rm T}$) can be 
summarized in the form of the matrix equation,
\newpage
\begin{widetext}
\begin{align}
\label{eq:langeem}
\mathcal{M}\left(\begin{array}{c}
x_{i}\\
\bm{\pi}_{\mathrm{L}}\\
\theta
\end{array}\right)\equiv\left(\begin{array}{rrrrr}
i\omega-A_{\sigma\sigma}-a_{\sigma\sigma}\bm{q}^{2}& -A_{\sigma n}-a_{\sigma n}\bm{q}^{2} &-A_{\sigma\varepsilon}-a_{\sigma\varepsilon}\bm{q}^{2}& -ia_{\sigma\pi}\bm{q}& 0\\
-a_{n\sigma}\bm{q}^{2} & i\omega-a_{nn}\bm{q}^{2}&  -a_{n\varepsilon}\bm{q}^{2}  &-ia_{n\pi}\bm{q}  &-a_{n\theta}\bm{q}^{2}\\
-a_{\varepsilon\sigma}\bm{q}^{2} & -a_{\varepsilon n}\bm{q}^{2} & i\omega-a_{\varepsilon\varepsilon}\bm{q}^{2} & -ia_{\varepsilon\pi}\bm{q}&  -a_{\varepsilon\theta}\bm{q}^{2}\\
-ia_{\pi\sigma}\bm{q} & -ia_{\pi n}\bm{q} & -ia_{\pi\varepsilon}\bm{q} & i\omega-a_{\pi\pi}\bm{q}^{2} & 0\\
-A_{\theta\sigma}-a_{\theta\sigma}\bm{q}^{2} &-A_{\theta n}-a_{\theta n}\bm{q}^{2} &-A_{\theta\varepsilon}-a_{\theta\varepsilon}\bm{q}^{2}& -ia_{\theta\pi}\bm{q} &i\omega-a_{\theta\theta}\bm{q}^{2}
\end{array}\right)\left(\begin{array}{c}
\sigma\\
 n\\
\varepsilon\\
\bm{\pi}_{\mathrm{L}}\\
\theta
\end{array}\right)=0\,,
\end{align}
\end{widetext}
where $A_{\alpha\beta}$ and $a_{\alpha\beta}$ are the parameters depending on $V_{ij},\ v_{ij}$, 
kinetic coefficients, and thermodynamic quantities. Note that $A_{\alpha\beta},\ a_{\alpha\beta}$ 
are not symmetric with respect to $\alpha$ and $\beta$. 

The eigenfrequencies of Eq.~(\ref{eq:langeem}) can be found from $\det\mathcal{M}=0$,
which yields
\begin{align}
\label{eq:proper}
\omega^{5}&+i\left(A_{\sigma\sigma}+g_{4}\bm{q}^{2}\right)\omega^{4}
-\left[g_{3}\bm{q}^{2}+O(\bm{q}^{4})\right]\omega^{3}\notag\\
&-i\left[g_{2}\bm{q}^{2}+O(\bm{q}^{4})\right]\omega^{2}+\left[g_{1}\bm{q}^{4}+O(\bm{q}^{6})
\right]\omega\notag\\
&\quad\quad\quad\quad\qquad\qquad\qquad+i\left[g_{0} \bm{q}^{4}+O(\bm{q}^{6})\right]=0\,,
\end{align}
where $g_0, \cdots, g_{4}$ are the functions of $A_{\alpha\beta}$ and $a_{\alpha\beta}$. 
Among others, we only give the explicit 
expressions for $A_{\sigma\sigma}$, $g_{2}$, and $g_{0}$,
\begin{align}
A_{\sigma\sigma}&\equiv \Gamma_{\sigma\sigma}\,,\\
\label{eq:zpara}
g_{2}&\equiv
\Gamma_{\sigma\sigma}
\left(n^{2}_{\text{eq}}V_{\pi\pi}+V_{\theta\theta}\right)\left|\begin{array}{cc}
V_{\sigma\sigma} &V_{\sigma n}\notag\\
V_{\sigma n} &V_{nn}
\end{array}\right|\\
&\quad+2\Gamma_{\sigma\sigma} \left(n_{\text{eq}}w_{\text{eq}}V_{\pi\pi}+\mu V_{\theta\theta}\right)\left|\begin{array}{cc}
V_{\sigma\sigma}&V_{\sigma n}\\
V_{\sigma\varepsilon}&V_{n\varepsilon}
\end{array}\right|\notag \\
&\quad+\Gamma_{\sigma\sigma} (w_{\text{eq}}^{2}V_{\pi\pi}+\mu^{2} V_{\theta\theta})\left|\begin{array}{cc}
V_{\sigma\sigma}&V_{\sigma\varepsilon}\\
V_{\sigma\varepsilon}&V_{\varepsilon\varepsilon}
\end{array}\right|\,,\\
\label{eq:psi}g_{0}
&\equiv \displaystyle \Gamma_{\sigma\sigma}V_{\pi\pi}V_{\theta\theta}T^{2} s_{\text{eq}}^{2} \det V\,,
\end{align}
which will be used in the following discussion.

Equation~(\ref{eq:proper}) can be factorized as
\begin{align}
&\left[\omega+iA_{\sigma\sigma}+i\left(g_{4}-\frac{g_{3}}{A_{\sigma\sigma}}+\frac{g_{2}}{A_{\sigma\sigma}^{2}}\right)\bm{q}^{2}\right]\notag \\
&\times\left[\omega-s_{+}|\bm{q}|+i\frac{t_{+}}{2}\bm{q}^{2}\right]\left[\omega+s_{+}|\bm{q}|+i\frac{t_{+}}{2}\bm{q}^{2}\right]\notag \\
\label{eq:proper2}&\times\left[\omega-s_{-}|\bm{q}|+i\frac{t_{-}}{2}\bm{q}^{2}\right]\left[\omega+s_{-}|\bm{q}|+i\frac{t_{-}}{2}\bm{q}^{2}\right]=0\,,
\end{align}
where $s_{\pm}$ and $t_{\pm}$ satisfy
\begin{align}
s_{+}^{2}+s_{-}^{2}&=\displaystyle \frac{g_{2}}{A_{\sigma\sigma}}\,,\\
s_{+}^{2}s_{-}^{2}&=\displaystyle \frac{g_{0}}{A_{\sigma\sigma}}\,,\\
t_{+}+t_{-}&=\displaystyle \frac{g_{3}}{A_{\sigma\sigma}}-\frac{g_{2}}{A_{\sigma\sigma}^{2}}\,,\\
s_{+}^{2}t_{-}+s_{-}^{2}t_{+}&=\displaystyle \frac{g_{1}}{A_{\sigma\sigma}}-\frac{g_{0}}{A_{\sigma\sigma}^{2}}\,.
\end{align}

From Eq.~(\ref{eq:proper2}), one finds all the hydrodynamic modes, except for the diffusion 
mode described by Eq.~(\ref{eq:pitdynamics}): one relaxation mode and two pairs of phonons. 
The speeds of phonons, $s_{\pm}$, can be obtained from the solution of
\begin{align} 
s^{4}-\frac{g_{2}}{A_{\sigma\sigma}}s^{2}+\frac{g_{0}}{A_{\sigma\sigma}}=0\,.
\end{align}
Note here that Eq.~(\ref{eq:psi}) shows that $g_{0} \rightarrow 0$ as the critical point
is approached, $\det V\rightarrow 0$. Near the critical point, we thus obtain 
\begin{align}
s_{+}^{2}=\displaystyle \frac{g_{2}}{A_{\sigma\sigma}}-\frac{g_{0}}{g_{2}}\,,\quad s_{-}^{2}=\frac{g_{0}}{g_{2}}\,.
\end{align}

\subsubsection{Dynamic critical exponent}
From the results above, we can show that one of the phonons with the speed $c_{\rm s}\equiv s_{-}$ 
exhibits the critical slowing down as follows. By using Eqs.~(\ref{eq:zpara}) and (\ref{eq:psi}), 
together with Eqs.~(\ref{eq:chiepep})--(\ref{eq:chinn}), we have
\begin{align}
\label{eq:cs2}
c_{\rm s}^{2}&=\displaystyle \frac{V_{\pi\pi}V_{\theta\theta}T^{3}s_{\mathrm{e}\mathrm{q}}^{2}}{\kappa_{nn}\chi_{nn}+2\kappa_{n\varepsilon}\chi_{n\varepsilon}+\kappa_{\varepsilon\varepsilon}
\chi_{\varepsilon\varepsilon}}\,,
\end{align}
where
\begin{align}
\kappa_{nn}&\equiv n_{\mathrm{e}\mathrm{q}}^{2}V_{\pi\pi}+V_{\theta\theta},\\
\kappa_{n\varepsilon}&\equiv n_{\mathrm{e}\mathrm{q}}w_{\mathrm{e}\mathrm{q}}V_{\pi\pi}+\mu V_{\theta\theta},\\
\kappa_{\varepsilon\varepsilon}&\equiv w_{\mathrm{e}\mathrm{q}}^{2}V_{\pi\pi}+\mu^{2}V_{\theta\theta}\,.
\end{align}
Using Eq.~(\ref{eq:chixis}), $V_{\pi\pi}\sim \xi^0$, and $V_{\theta\theta}\sim \xi^0$ 
near the critical point, we obtain
\begin{align}
c_{\rm s}^2 \sim \xi^{-2 + \eta}\,.
\end{align} 
Comparing it with Eq.~(\ref{eq:zdef}), we find that the dynamic critical exponent $z$ is again 
given by Eq.~(\ref{eq:z}): the dynamic universality class remains the same as the case without 
the coupling to $\varepsilon$ and ${\bm \pi}$ in Sec.~\ref{sec:dynamics}. 

Finally, we remark that the dynamic critical exponent $z$ obtained above is not affected by 
nonlinear couplings beyond the mean field. This can be understood by recalling that what is 
renormalized by nonlinear terms is the kinetic coefficient \cite{Chaikin} and that Eq.~(\ref{eq:cs2}) 
does not depend on any kinetic coefficient. This should be contrasted with the case of the 
high-temperature QCD critical point, where the dynamic critical exponent is modified by 
nonlinear couplings \cite{Son:2004iv}. This is because the hydrodynamic mode that exhibits the 
critical slowing down there is the diffusion mode, whose diffusion rate depends on a kinetic 
coefficient.

\section{Conclusion and discussion}
\label{sec:conclusion}
In this paper, we have shown that the high-density QCD critical point belongs to the new dynamic 
universality class that is beyond the conventional classification by Hohenberg and Halperin. 
We have demonstrated that the speed of the superfluid phonon vanishes as the critical point is 
approached and that the dynamic critical index is $z \approx 2$. 

The physical reason why the dynamic universality class of the high-density QCD critical point is 
new can be understood as follows. First, it is different from that of the high-temperature QCD 
critical point (i.e., model H \cite{Son:2004iv}) due to the presence of the superfluid phonon 
associated with the $\U(1)_{\rm B}$ symmetry breaking. Second, it must also be different 
from that of the superfluid transition of $^{4}$He despite the presence of superfluid phonons in 
both cases. This is because the former criticality is characterized by the vanishing chiral order 
parameter (with a nonzero superfluid gap), while the latter is characterized by the vanishing 
superfluid gap (or vanishing stiffness constant).%
\footnote{In fact, this difference of order parameters is reflected in the difference of the critical 
behaviors of the speeds of superfluid phonons between the two cases. In the superfluid $^{4}$He, 
the speed of superfluid phonon is given by 
$c_{\rm s}=\sqrt{\rho_{\rm s}/{c_{\rm p}}}$ \cite{Hohenberg:1977ym}, where $\rho_{\rm s}$ is 
the stiffness constant and $c_{\rm p}$ is the specific heat at constant pressure. When the 
superfluid transition is approached (i.e., when the correlation length diverges, $\xi \rightarrow \infty$), 
the stiffness constant goes to zero as $\rho_{\rm s}\sim\xi^{-1}$ and $c_{\rm p}\sim\xi^{\alpha/\nu}$. 
[Here, $\alpha$ and $\nu$ are the critical exponents of $c_{\rm p}$ and $\xi$, defined as 
$c_{\rm p}\sim\tau^{-\alpha}$ and $\xi\sim\tau^{-\nu}$ with $\tau\equiv(T-T_{\rm c})/T_{\rm c}$.]
In particular, the $\xi$ dependence of $c_{\rm s}$ is different from Eq.~(\ref{eq:cs}), because 
the high-density QCD critical point is away from the superfluid transition and the stiffness 
parameter $d$ in Eq.~(\ref{eq:GL}) remains {\it nonvanishing}.}

Our results suggest that, while the static critical phenomena {\it cannot} distinguish between 
the high-temperature and high-density QCD critical points, the dynamic critical phenomena 
{\it can} distinguish between the two. 
It would be important to study possible phenomenological consequences and experimental 
signatures of the high-density critical point to be tested in future heavy ion collision experiments, 
similarly to the high-temperature one \cite{Stephanov:1998dy}. 
The uniqueness of the dynamic critical phenomena around the high-density QCD critical point, 
if observed, would provide indirect evidence of the superfluidity in high-density QCD matter.

Finally, it would also be interesting to study the possible effects of dynamical electromagnetic fields
(or massless photons), which may affect not only the dynamic universality class of the high-density 
critical point, but also even that of the high-temperature one.

\acknowledgments
We thank H.~Fujii and Y.~Hidaka for useful discussions. 
This work was supported by JSPS KAKENHI Grant No.~16K17703 and MEXT-Supported Program 
for the Strategic Research Foundation at Private Universities, ``Topological Science" (Grant No.~S1511006).
N.~S. acknowledges the Research Grant of Keio Leading-edge Laboratory of Science and 
Technology and the fellowship of the European Physical Society (EPS) provided through 
the International School of Nuclear Physics 2016 in Erice, where this work was completed.


\begin{thebibliography}{99}

  \bibitem{Fukushima:2010bq} 
  K.~Fukushima and T.~Hatsuda,
  Rep.\ Prog.\ Phys.\  {\bf 74}, 014001 (2011).
  
  \bibitem{Stephanov:2004wx} 
  M.~A.~Stephanov,
  Prog.\ Theor.\ Phys.\ Suppl.\  {\bf 153}, 139 (2004)
  [Int.\ J.\ Mod.\ Phys.\ A {\bf 20}, 4387 (2005)].

  \bibitem{Asakawa:1989bq}
  M.~Asakawa and K.~Yazaki,
  Nucl.\ Phys.\ {A\bf 504}, 668 (1989).
  
  \bibitem{Barducci:1989wi}
  A.~Barducci, R.~Casalbuoni, S.~De Curtis, R.~Gatto, and G.~Pettini,
  Phys.\ Lett.\ B {\bf 231}, 463 (1989).
  
  \bibitem{Halasz:1998qr} 
  A.~M.~Halasz, A.~D.~Jackson, R.~E.~Shrock, M.~A.~Stephanov, and J.~J.~M.~Verbaarschot,
  Phys.\ Rev.\ D {\bf 58}, 096007 (1998).
  
  \bibitem{Berges:1998rc} 
  J.~Berges and K.~Rajagopal,
  Nucl.\ Phys.\ {\bf B538}, 215 (1999).
  
  \bibitem{Hatta:2002sj} 
  Y.~Hatta and T.~Ikeda,
  Phys.\ Rev.\ D {\bf 67}, 014028 (2003).
  
  \bibitem{Hatsuda:2006ps}
  T.~Hatsuda, M.~Tachibana, N.~Yamamoto, and G.~Baym,
  Phys.\ Rev.\ Lett.\  {\bf 97}, 122001 (2006); 
  Phys.\ Rev.\ D {\bf 76}, 074001 (2007).
  
  \bibitem{Abuki:2010jq}
  H.~Abuki, G.~Baym, T.~Hatsuda, and N.~Yamamoto,
  Phys.\ Rev.\ D {\bf 81}, 125010 (2010).
  
  \bibitem{Schmitt:2010pf} 
  A.~Schmitt, S.~Stetina, and M.~Tachibana,
  Phys.\ Rev.\ D {\bf 83}, 045008 (2011).
    
  \bibitem{Kitazawa:2002bc}
  M.~Kitazawa, T.~Koide, T.~Kunihiro, and Y.~Nemoto,
  Prog.\ Theor.\ Phys.\  {\bf 108}, 929 (2002).
  
  \bibitem{Alford:1998mk} 
  M.~G.~Alford, K.~Rajagopal, and F.~Wilczek,
  Nucl.\ Phys.\ {\bf B537}, 443 (1999).
  
  \bibitem{Fujii:2003bz} 
  H.~Fujii,
  Phys.\ Rev.\ D {\bf 67}, 094018 (2003).
  
  \bibitem{Fujii:2004jt} 
  H.~Fujii and M.~Ohtani,
  Phys.\ Rev.\ D {\bf 70}, 014016 (2004).
  
  \bibitem{Son:2004iv} 
  D.~T.~Son and M.~A.~Stephanov,
  Phys.\ Rev.\ D {\bf 70}, 056001 (2004).
  
  \bibitem{Minami:2011un} 
  Y.~Minami,
  Phys.\ Rev.\ D {\bf 83}, 094019 (2011).          

  \bibitem{Hohenberg:1977ym} 
  P.~C.~Hohenberg and B.~I.~Halperin,
  Rev.\ Mod.\ Phys.\  {\bf 49}, 435 (1977).
  
  \bibitem{Alford:2007xm} 
  M.~G.~Alford, A.~Schmitt, K.~Rajagopal, and T.~Sch\"{a}fer,
  Rev.\ Mod.\ Phys.\  {\bf 80}, 1455 (2008).
    
  \bibitem{Chaikin}
  P.~M.~Chaikin and T.~C.~Lubensky,
  {\it Principles of Condensed Matter Physics}
  (Cambridge University Press, Cambridge, England, 1995).

  \bibitem{Weinberg:1996kr} 
  S.~Weinberg,
  {\it The Quantum Theory of Fields}, Modern applications Vol. 2 (Cambridge University Press, Cambridge, England, 1996).
  
  \bibitem{Son:2002zn} 
  D.~T.~Son,
  hep-ph/0204199.                         
  
  \bibitem{Volovick}
  I.~E.~Dzyaloshinskii and G.~E.~Volovick,
  Ann. Phys. (N.Y.)\  {\bf 125}, 67 (1980).
  
  \bibitem{Stephanov:1998dy} 
  M.~A.~Stephanov, K.~Rajagopal, and E.~V.~Shuryak,
  Phys.\ Rev.\ Lett.\  {\bf 81}, 4816 (1998);
  Phys.\ Rev.\ D {\bf 60}, 114028 (1999).

\end{thebibliography}
\end{document}